\documentclass{article}

\usepackage{microtype}
\usepackage{graphicx}
\usepackage{subfigure}
\usepackage{booktabs}

\usepackage{amsmath, amssymb}
\usepackage{multicol, multirow}
\usepackage{xspace}
\usepackage{xcolor}

\usepackage{hyperref}

\usepackage[accepted]{mlsys2025style/mlsys2025}

\mlsystitlerunning{Awesome Paper}

\begin{document}

\twocolumn[
\mlsystitle{Trellis: Bringing Compiler-Guided Repair into Agentic Code Generation}

\begin{mlsysauthorlist}
\mlsysauthor{Alexander Du}{Duke}
\mlsysauthor{Jianjun Ou}{Duke}
\mlsysauthor{Danyang Zhuo}{Duke}
\mlsysauthor{Matthew Lentz}{Duke}
\end{mlsysauthorlist}

\mlsysaffiliation{Duke}{Department of Computer Science, Duke University, Durham, NC, USA}
\mlsyscorrespondingauthor{Alexander Du}{alexander.du@duke.edu}
\mlsyskeywords{Awesome Keywords}

\vskip 0.3in

\newcommand{\sys}{Trellis\xspace}
\newcommand{\tsc}{\texttt{tsc}\xspace}
\begin{abstract}

Coding agents regularly generate programs that do not compile and require substantial tokens to repair.
Yet most repairs are small and local to the edit that caused them.
We present \sys, a system that integrates incremental compiler feedback into code generation to improve code correctness while controlling repair costs.
\sys uses rollback-based search for targeted repair and asynchronous checking and checkpointing for efficient execution.
We adapt Clang and \tsc to \sys's checker interface and integrate \sys transparently into OpenCode.
On C/C++ and TypeScript software engineering tasks, \sys improves compilation rate by up to 13\% and test pass rate by up to 11\% with comparable token consumption to baseline OpenCode.

\end{abstract}

]

\printAffiliationsAndNotice{}

\newcommand{\sys}{Trellis\xspace}
\newcommand{\tsc}{\texttt{tsc}\xspace}
\section{Introduction}
\label{sec:introduction}

Coding agents, such as Codex~\cite{openai2026codex}, Claude Code~\cite{anthropic2026claudecode}, OpenCode~\cite{opencode}, and OpenHands~\cite{openhands}, are rapidly changing how developers write, test, and review software.
With access to compilers, tests, and other tools, agents can check and revise generated code over successive interaction turns~\cite{chen2024teaching, yang2024sweagent, openhands}.

Tool access, however, does not ensure prompt error detection or efficient repair.
Unchecked errors can affect subsequent generation or remain unresolved when an agent considers a task complete~\cite{selfconditioning, lostmultiturn}.
Even when errors are caught, agents spend additional turns repairing problems that may admit localized fixes.
Repeated cycles of reasoning, checking, and revision consume time and tokens, increasing cost to the user~\cite{olausson2024selfrepair, agentdiet}.

Compiler errors are a useful target for localized checking and repair: they prevent generated programs from being tested or executed, and compilers provide well-defined rules for detecting them.
In our evaluation of several open-weight models, compiler-error repair accounts for as much as 27\% of token use and elapsed time across attempts involving repair.
Prior work on constrained semantic decoding~\cite{poesia2022synchromesh, mundler2025typeconstrained, li2025correctness, nagy2026chopchop, li2025adaptrack} and incremental compiler feedback~\cite{agrawal2023monitorguided, wei2023copiloting, jiang2025rocode} demonstrates the promise of checking and repairing code during standalone LLM code generation.
However, these approaches rely on restricted~\cite{poesia2022synchromesh, mundler2025typeconstrained} or approximate~\cite{agrawal2023monitorguided, wei2023copiloting} checking, leave checking and repair costs unaddressed, and do not accommodate agent workflows.

\begin{figure}[t]
    \centering
    \includegraphics[width=\linewidth]{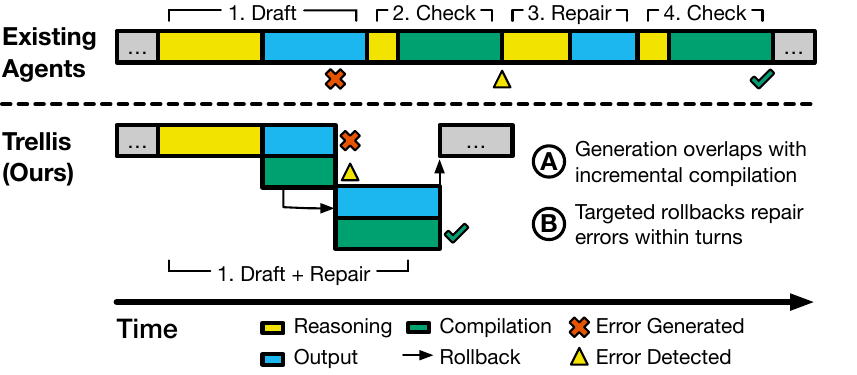}
    \vskip -0.1in
    \caption{Agent-directed compiler-error repair requires additional turns to request checks and revise code. \sys overlaps compilation with generation and initiates localized repair within a generation turn.}
    \label{fig:intro}
    \vskip -0.1in
\end{figure}

We present \sys, a system that brings compiler-guided repair into agentic code generation.
Our goal is to improve the correctness of generated code with comparable token consumption.
\sys intercepts file-writing tool calls and streams the generated code to a compiler, overlapping checking with generation (Figure~\ref{fig:intro}).
When the compiler reports an error, \sys attempts a localized repair by restarting generation from an earlier code prefix.
It then returns control to the agent, providing diagnostic feedback if errors remain.

Realizing this approach requires overcoming three challenges.
First, production compilers such as Clang and \tsc do not natively provide sound diagnostics on partial input streams.
Adapting them requires reusing analyses designed for complete program constructs while providing timely feedback on unfinished code.
%
Second, local repair must improve correctness while keeping checking and repair overhead manageable.
This requires choosing economical repair points, reusing computation across repair attempts, and deciding when further retries are worth less than returning control to the agent.
Third, local repair must accommodate intermediate steps in the agent's broader workflow.
An error may require subsequent edits across files or turns to resolve, so requiring every edit to compile can obstruct the intended change.

We adapt the Clang \cite{llvm2026clang} and \tsc \cite{microsoft2026typescript} compilers to support \sys and integrate it into OpenCode.
On C/C++ and TypeScript software engineering tasks \cite{jimenez2024swebench, zan2025multiswebench}, \sys-backed OpenCode increases compilation rate by up to 13\% and test pass rate by up to 11\%, with token consumption comparable to baseline OpenCode.

In summary, we make the following contributions:
\begin{enumerate}

\item An incremental compiler interface and design for sound diagnostics on partially generated code, instantiated in Clang and \tsc.

\item A system for efficient checking and repair, combining asynchronous checking, reusable model and compiler checkpoints, and rollback-based search.

\item An integration that coordinates local repair with subsequent agent actions, evaluated in OpenCode on C/C++ and TypeScript software engineering tasks.

\end{enumerate}
\section{Motivation}
\label{sec:motivation}

Coding agents operate through repeated interactions between an agent harness and an LLM.
The harness exposes tools to the model; for example, OpenCode provides the \texttt{write} and \texttt{edit} tools for modifying code and a shell tool for general commands.
Each turn, the model generates a response that may include reasoning, exposition, and tool calls.
The harness executes the calls and adds their results to the context for subsequent turns.

Checking and repair depend on the model choosing when to invoke tools, interpreting their results, and issuing subsequent edits.
This interaction loop can delay error detection and consume substantial resources even when the required fix is small.
In this section, we examine the prevalence of compiler errors, the cost of agent-directed repair, and the opportunity for local repair within generation turns.

\paragraph{Error prevalence and detection.}
To understand the prevalence of compiler errors, we evaluated three recent, open-weight models---Nemotron 3 30B (\emph{Nano})~\cite{nvidia2025nemotron3}, 120B (\emph{Super}), and Gemma 4 31B (\emph{Gemma})~\cite{gemma4}---on C/C++ and TypeScript software engineering tasks using OpenCode.
Our evaluation setup is detailed in \S\ref{sec:evaluation}.
%

Out of attempts that produce a final patch, 24--37\% introduce at least one compiler error.
Of these, 30--74\% end with compiler errors in the final patch.
Moreover, 17--38\% of attempts ending with compiler errors contain at least one unchecked error: the agent never compiles the affected file after introducing it.

\begin{figure}[t!]
    \centering
    \includegraphics[width=\linewidth]{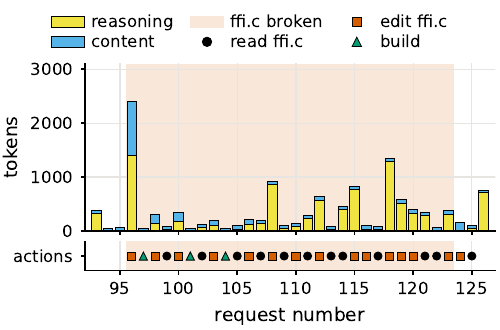}
    \vskip -0.1in
    \caption{A repair window from a Super execution trace, showing generated tokens and tool calls.}
    \label{fig:motivation-trace}
    \vskip -0.1in
\end{figure}

\begin{table}[t!]
    \caption{Output tokens and latency attributed to repair windows.  Absolute values are means per window; percentages are shares of total usage across affected attempts.  Overlapping windows are counted only once.}
    \label{tab:motivation-cost}
    \vskip 0.1in
    \begin{center}
    \begin{small}
    \begin{sc}
        \begin{tabular}{lccc}
        \toprule
            Model & Nano & Super & Gemma \\
        \midrule
            \# & 87 & 166 & 98 \\
            Tok (k) & 11.5 (26.5\%) & 5.2 (15.3\%) & 1.9 (14.6\%) \\
            Lat (s) & 160 (26.7\%) & 207 (16.5\%) & 93 (16.7\%) \\
        \bottomrule
        \end{tabular}
    \end{sc}
    \end{small}
    \end{center}
    \vskip -0.1in
\end{table}

\paragraph{Repair cost.}
Even successful repair can consume substantial resources.
We define a repair window as a sequence of actions beginning with an edit that introduces a compiler error and ending when the file compiles again.
Figure~\ref{fig:motivation-trace} illustrates one such window from a Super execution trace.
The window (shaded) begins with an edit that breaks \texttt{ffi.c} and ends when the file compiles again.
During this window, the agent repeatedly reads, reasons about, and edits the file, generating around 8.5k tokens over 300s.

Table~\ref{tab:motivation-cost} quantifies resource usage across repair windows.
Costs are difficult to attribute precisely---agents may interleave repair with other work, and a reasoning step may serve multiple purposes.
Therefore, we take a conservative approach by considering only windows in which an error is detected and repaired, counting only resources consumed after detection, and including only turns that directly interact with the broken file.
This excludes unsuccessful repairs and omits repair work involving other files.
Nonetheless, the attributed costs are substantial: the average repair window accounts for thousands of generated tokens and minutes of elapsed time.
Across attempts containing a repair window, attributed costs account for 15\%--27\% of total usage, depending on the model and resource measured.

\paragraph{Repair locality.}

\begin{figure}[t!]
    \centering
    \includegraphics[width=0.9\linewidth]{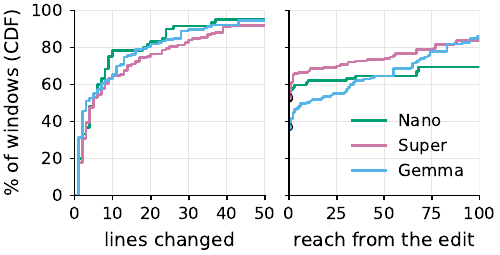}
    \vskip -0.1in
    \caption{Size and locality of observed repairs.  Left: lines changed between the broken and repaired file states.  Right: fraction of repair windows in which retaining only changes within the original edit window and a surrounding margin still repairs the compiler error.}
    \label{fig:motivation-locality}
    \vskip -0.1in
\end{figure}

Given the cost of agent-directed repair, could some compiler errors be fixed within the same generation turn?
For each repair window, we compare the broken and repaired file states to measure the size of the repair and its extent beyond the edit that introduced the error.
Small repairs confined to the original edit suggest an opportunity to fix errors within the same generation turn, avoiding further agent interactions.

Figure~\ref{fig:motivation-locality} shows that observed repairs are often small and local.
Over half change only one to five lines of code.
For 40--60\% of successful repairs, retaining only changes within the original edit window is sufficient to repair the compiler error.
Allowing changes in the surrounding margins increases this fraction.
These results represent a lower bound on opportunities for local repair, since errors repaired elsewhere may also admit local fixes.

These results motivate selective repair.
Many observed fixes are small and confined to the original edit, despite substantial agent-directed repair costs.
A system should exploit inexpensive local fixes while preserving the agent's ability to make broader changes across edits and files.
\section{Overview}
\label{sec:overview}

\begin{figure}[t!]
    \centering
    \includegraphics[width=\linewidth]{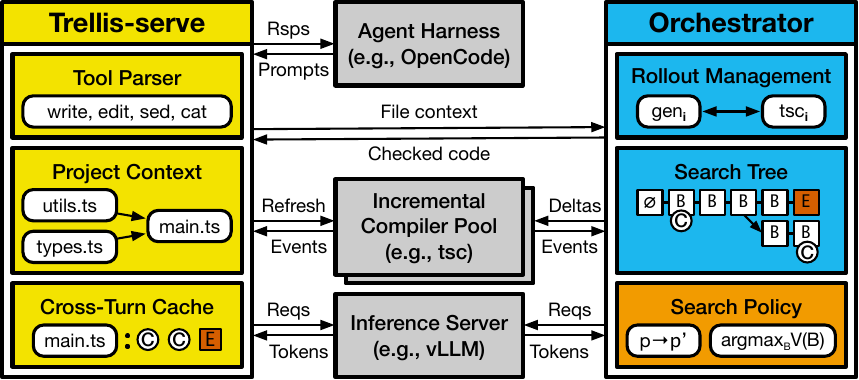}
    \vskip -0.1in
    \caption{\sys architecture components.}
    \label{fig:overview-architecture}
    \vskip -0.1in
\end{figure}

\sys checks generated code and attempts localized repairs within an agent's generation turn.
It integrates transparently through the agent's existing inference interface, bringing compiler-guided repair to supported file-writing operations. 
Figure~\ref{fig:overview-architecture} presents an overview of the \sys architecture.

\sys-serve forwards ordinary generation unchanged, but intercepts supported file-writing operations as the model begins producing code.
It supplies the surrounding file context to the orchestrator, which pairs generation with an incremental compiler session to form a \emph{rollout}.
The compiler identifies semantically meaningful resume points, exposes reusable checkpoints, and reports errors once they cannot be resolved by additional input.

The orchestrator overlaps compilation with generation.
It organizes compiler-identified resume points into a search tree and queries a repair policy to choose where to restart generation when an error arrives.
The policy interface supports diverse search strategies.
Our chosen policy uses Bayesian estimates to select restarts with the lowest expected repair cost.

To support changes across edits and turns, \sys tracks pre-existing errors and tolerates selected errors that may result from changes in progress.
When local repair becomes unproductive, it returns control to the agent with diagnostic feedback.
\sys-serve maintains the evolving file context and retains compiler checkpoints and diagnostics across turns, invalidating and refreshing them when relevant source files or dependencies change.

\begin{figure}[t!]
    \centering
    \includegraphics[width=\linewidth]{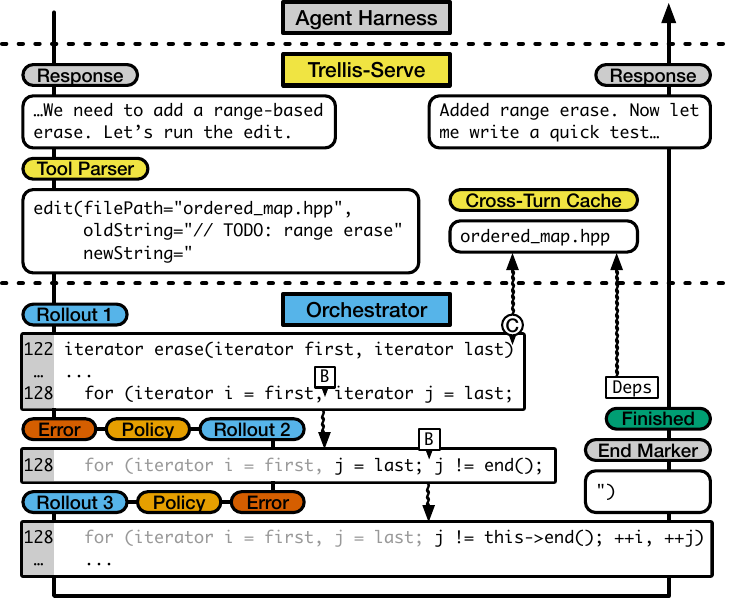}
    \vskip -0.1in
    \caption{\sys execution walkthrough.  The execution path starts at the top left of the figure and ends at the top right.}
    \label{fig:overview-walkthrough}
    \vskip -0.1in
\end{figure}

\paragraph{Walkthrough.}
Figure~\ref{fig:overview-walkthrough} illustrates \sys's workflow using an edit inspired by a Super execution trace.
The model intends to add a range-based erase operation to \texttt{ordered\_map.hpp}.
Before generation of \texttt{newString} begins, \sys-serve identifies the target file, matches \texttt{oldString} to locate the edit region, and passes control of generation to the orchestrator.

The orchestrator pairs the generator session with a compiler session to form the first rollout.
While generation begins from the edit, the compiler checks from the start of the file: the orchestrator first feeds the compiler the unchanged file prefix, then streams generated tokens as they arrive.
As the rollout progresses, the compiler emits boundary events marking semantically meaningful resume points.
Some boundaries also carry compiler checkpoints.
The figure shows a boundary and checkpoint after the function declaration (Rollout 1, Line 122) and two boundaries within the loop (Rollouts 1 and 2, Line 128); other boundaries are omitted for clarity.

The rollout stumbles: the model repeats \texttt{iterator} in the loop initializer (Rollout 1, Line 128).
The compiler reports an error, which the orchestrator forwards to the policy.
The policy updates its repair-success probability estimates and selects the boundary with the lowest expected repair cost.

The orchestrator restarts generation at the chosen boundary.
Compilation resumes from the closest preceding checkpoint, here after the function declaration, and replays the intervening text.
This allows generation to restart within the loop without recompiling the entire file prefix.

This process repeats until the model closes the \texttt{newString} argument with the \texttt{edit} call's end marker.
The orchestrator feeds the compiler the unchanged file suffix and waits for the final verdict.

The edit succeeded!
\sys-serve updates its project context and cross-turn cache with the compiler-reported dependencies and checkpoints forked during execution.
The orchestrator returns control of the generator session to \sys-serve, and the model continues its response unconstrained.

\section{Design}
\label{sec:design}

\sys's design addresses three challenges: obtaining useful feedback from production compilers during generation, making checking and repair economical, and cooperating with the agent's broader workflow.

\subsection{Adapting Production Compilers}
\label{sec:design-adapting}

Checking partially generated programs requires distinguishing real errors from code that is merely unfinished.
Most production compilers do not natively support this distinction on streamed input.
Existing approaches use specialized checkers~\cite{mundler2025typeconstrained}, language server queries~\cite{agrawal2023monitorguided}, or tools to complete partial programs such that they can be input to standard compilers \cite{jiang2025rocode}.
These approaches offer limited coverage of language features or compiler checks (in part due to substantial implementation effort), and none fully address the overhead of repeatedly checking shared prefixes.

We instead explore a different point in the design space by \emph{adapting compiler front-ends to consume streamed input and reuse their existing analyses.}
One key observation drives our approach: the compiler need not report a verdict after every token; it can report results as the relevant analyses complete.
This matches how production compilers operate: analyses work on program constructs, and some diagnostics depend on later declarations or uses.
Generation proceeds asynchronously, allowing the compiler to defer diagnostics until it has enough input to report them soundly.

\paragraph{Interface.}
\sys communicates with incremental compilers through a small streaming interface.
The orchestrator streams program text to the compiler; closing the stream signals the end of input.
The compiler emits \emph{boundary}, \emph{error}, and \emph{finished} events.
Boundary and error events identify positions in the input, allowing the orchestrator to relate compiler feedback to generated prefixes.

\paragraph{Boundaries and checkpoints.}
Boundaries mark semantically meaningful places where \sys can restart generation, such as the ends of statements and declarations.
Each boundary event includes a character offset and may carry metadata (e.g., kind, brace depth).
These boundaries give the repair policy a discrete set of candidate prefixes aligned with program structure.
A boundary provides no correctness guarantee, since analyses of the preceding code may still be pending.

A checkpoint optionally associates a boundary with saved compiler state.
We require checkpoints to be \emph{resumable}: continuing from a checkpoint with a replacement suffix must be equivalent to compiling the preserved prefix and replacement suffix from scratch under the same project context.
To restart from a boundary without a checkpoint, the orchestrator restores the nearest earlier checkpoint and replays the intervening text, or starts from scratch if none are available.

\paragraph{Errors.}
An error event carries a diagnostic and a \emph{limit}: the latest character position from which a replacement continuation can still avoid the error.
The limit is often the diagnostic's reported location, but the two can differ.
For example, an unused-name diagnostic points to a definition but can be prevented by inserting a use later in the enclosing scope.
Error events may also carry metadata, such as an error code or compiler fix-its.

To avoid rejecting valid continuations, we require reported errors to be \emph{final}: extending the current input cannot make them disappear.
A diagnostic that depends on an unfinished construct must therefore be delayed until the compiler can establish finality.
This delay introduces \emph{diagnostic latency}, the gap between generating an error and reporting it.

\paragraph{Completion and error recovery.}
Closing the input stream allows the compiler to complete deferred analyses.
Once checking is complete, the finished event reports the full error list and dependencies.
We require the incremental compiler to match the standard compiler's acceptance decision and diagnostic list under the same project context.
Dependencies identify when changes elsewhere in the project could invalidate the result.

Agent workflows may leave existing errors for subsequent edits to resolve.
We therefore also require \emph{resilience}: earlier errors must not compromise the validity of later diagnostics.
Syntax errors can violate this requirement by disrupting subsequent parsing.
We describe their treatment below.

\subsection{Efficient Checking and Repair}

With our incremental compiler design, we could simply overlap checking with generation and return feedback to the agent for revision in subsequent turns.
However, Figure~\ref{fig:motivation-locality} shows that many observed repairs fit within the original edit region.
We therefore explore how to efficiently search for local repairs before returning control to the agent.
We first discuss how the orchestrator executes rollouts and maintains a search tree, then explain how the policy chooses repairs and decides when further search is unlikely to succeed.

\paragraph{Rollouts and search tree management.}
The orchestrator pairs a compiler session with a generator session to form a \emph{rollout}, the fundamental unit of search in \sys.
Each rollout is defined by its starting code prefix and sampling parameters.
The orchestrator streams generated code to the compiler and processes compiler events as they arrive.
Generation proceeds without waiting for each check, so an error may be reported after generation has advanced beyond its cause.

For each retry, the orchestrator starts a generation request from a selected boundary and resumes compilation from the nearest preceding checkpoint, replaying any intervening text.
Reusable generator and compiler checkpoints reduce repeated computation: the underlying inference engine supplies generator checkpoints, while the incremental compiler supplies compiler checkpoints.
When a rollout completes the edit, the orchestrator feeds the compiler the unchanged file suffix and waits for the final verdict before deciding whether to accept it.

\begin{figure}
    \centering
    \includegraphics[width=\linewidth]{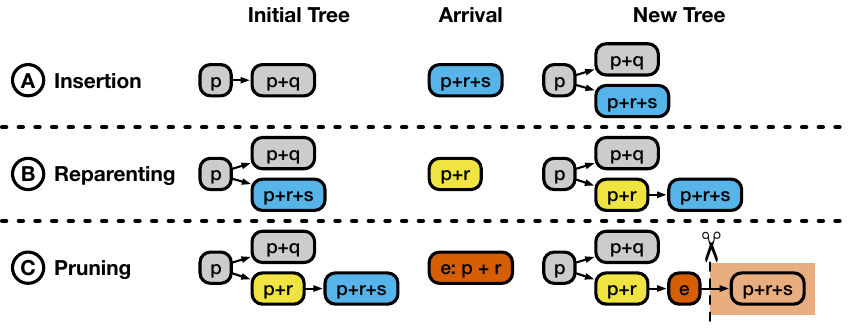}
    \vskip -0.1in
    \caption{Search-tree management in \sys.}
    \label{fig:design}
    \vskip -0.1in
\end{figure}

\sys represents explored continuations as a tree of code prefixes, with each rollout forging a path through the tree.
Each boundary event identifies a prefix and a candidate restart point.
The orchestrator de-duplicates boundaries by code hash and attaches each to its closest ancestor boundary.
The tree follows prefix relationships rather than event arrival order: if a boundary arrives out of order, the orchestrator reparents affected nodes to preserve these relationships.

An error event identifies a resume limit, which may precede boundaries already reported on its path. 
Continuations starting beyond that limit cannot repair the error.
If an error is not tolerated, the orchestrator inserts the error at its limit and prunes descendant resume points on that path.
Figure~\ref{fig:design} illustrates insertion, reparenting, and pruning.

\paragraph{Policy.}
\sys supports expressive policies that choose where to resume, how often to retry, and when to stop.
Whenever a boundary or error node is added, the orchestrator queries the policy with the event, a read-only view of the tree, the active rollouts, and the system budgets.
The policy can spawn rollouts from selected boundaries with chosen sampling parameters, kill rollouts, accept a rollout, or give up and return control to the agent.

In this paper, we choose a policy that makes decisions based on the cost of repair.
For each candidate boundary, it estimates the token cost and probability of a successful attempt~\cite{bellman1954theory}.
We also estimate the cost of giving up (i.e., returning control back to the agent) using repair-window analysis similar to that in \S\ref{sec:motivation}.
The policy continues local search only when its expected cost is lower than that of returning control to the agent.

The success probabilities are estimated by a Bayesian logistic regression with search-level and boundary-level features and random effects.
A failed attempt lowers the estimated success probability at that boundary and, more mildly, elsewhere in the search.
Repeated failures therefore encourage trying other boundaries or giving up.
We fit the model using offline sweeps and live \sys traces from the training set.
Appendix~\ref{app:policy} provides the formulation and estimation details.

To prevent repeated failures, our policy modifies retry generation in two ways.
First, it inserts a code comment at the resume point warning about the most recent error.
Second, it masks the final token of known-bad continuations to prevent their recurrence.
These bans are derived from observed failures and do not require enumerating all valid next tokens.

\paragraph{Returning control with feedback.}
When localized repair is not effective or economical, the policy returns control to the agent.
Overlapping checking and generation allows us to provide diagnostic feedback that the agent can use in subsequent turns, even when local repair gives up.
When the policy returns control, the orchestrator passes through a continuation sampled from the first rollout and reports remaining compiler errors in the tool-call response.

Syntax errors are a special case because they can corrupt subsequent diagnostics.
If an edit passed through after giving up introduces a syntax error, the edit is not applied.

\subsection{Agent Integration}

Agentic coding extends beyond generating a single source file.
An agent may edit several files across turns, issue multiple edits within one response, and modify dependencies through tools that \sys does not intercept.
Moreover, a change may require several edits before the project compiles again.
Supporting these workflows requires \sys to maintain the current project context for each edit and accommodate intermediate errors while the agent completes a broader change.

\paragraph{Maintaining project context.}
Checking an edit requires the current contents of both the target file and its dependencies, as well as knowledge of existing errors.
The agent harness applies edits between interaction turns, but the model may generate several edits in one response.
Checking each successive edit requires the file state produced by applying the preceding edits in order.
\sys-serve therefore maintains an overlay of files changed within the turn, making these temporary states available as edit targets and dependencies.

Files may also change through operations that \sys does not intercept, and the harness's applied edits may differ from the overlay.
At the start of each turn, \sys-serve reconciles the overlay with disk to prevent divergence from persisting across turns.

For each file, \sys-serve caches its \emph{status}---whether it is clean and if not, its error list---and any compiler checkpoints.
The error list supports matching pre-existing diagnostics, while checkpoints reduce compilation work on later edits.

\paragraph{Reusing compiler state.}
Checkpoints may be expensive, so \sys retains a bounded set per file.
Edits are highly clustered: the median distance between edits is a few hundred bytes, whereas the median edited file is tens of kilobytes.
We therefore retain more checkpoints near the edit and spread the remainder evenly across the file.
\sys-serve also bounds the number of cached files and evicts them by least recent use.
In practice, 99\% of edits target one of the last six touched files.

\paragraph{Keeping context current.}
Cached results depend on both source text and project context.
Changing an include file or creating a previously missing include can alter results even when the target file is unchanged.
\sys therefore tracks files the compiler read and paths it checked but did not find.
Cache entries also record a code hash: a status hashes the entire file, while a checkpoint hashes its code prefix.

At the start of each turn, \sys-serve checks cached files and their dependencies against disk.
If a file's modification time and size are unchanged, it reuses the cached hash; otherwise, it rehashes the file.
A changed hash marks affected statuses and checkpoints, including those of dependent files, as stale.

\sys-serve \emph{refreshes} a stale or missing cache entry by compiling the file's current version and observing compiler events.
It eagerly refreshes recently read or previously edited files, since these are likely to be edited again.
Other stale files are refreshed lazily when edited.
\sys-serve bounds concurrent refreshes and gives rollouts priority over background refreshes in the compiler pool.

\paragraph{Accommodating changes in progress.}
The project context also helps \sys distinguish errors introduced by an edit from those left by earlier actions.
Some errors require non-local fixes or reflect changes still in progress: for example, the agent may update a use site before adding the declaration that it depends on.
Insisting on a local repair can interfere with this intended sequence of edits.
\sys therefore allows the policy to decide which errors warrant repair and which to tolerate.

Our policy uses criteria based on error code, name, and position.
It tolerates pre-existing errors by matching against \sys-serve's cached error list.
It also tolerates errors outside the edit window involving names that occur within the edited window, since these may be consequences of an ongoing change.
\newcommand{\fork}{\texttt{fork()}}

\section{Implementation}
\label{sec:implementation}

We implement the core \sys components, \sys-serve and the orchestrator, in Python.
In this section, we describe how we intercept agent edits and adapt production compilers to support incremental checking and checkpointing.

\subsection{Intercepting Agent Edits}

\paragraph{Edits via Tool Calls.}
\sys-serve must identify file context from partial tool calls.
For example, OpenCode's edit tool takes the following arguments: \texttt{edit(filePath, oldString, newString, replaceAll)}.
Before \texttt{newString} is generated, \sys-serve identifies the file from \texttt{filePath} and the edit region from \texttt{oldString}.
If either argument follows \texttt{newString}, \sys lacks the context to check the edit.
Gemma's chat template sorts tool arguments alphabetically, placing \texttt{newString} before \texttt{oldString}; we therefore rename tool arguments for Gemma only.

\paragraph{Edits via Shell Commands.}
Models also modify files through shell commands.
\sys-serve intercepts two forms with explicit targets and replacement text: heredocs with quoted delimiters and \texttt{sed -i} with an \texttt{a}, \texttt{i}, or \texttt{c} command.
Other modifications pass through unchecked, including \texttt{sed -i 's/old/new/'}, \texttt{git} commands, and Python scripts that compute replacement text.
These unchecked modifications account for about 5--10\% of file modifications.

\subsection{Adapting Production Compilers (Clang, \tsc)}

Adapting Clang and \tsc required modifying approximately 3.5k lines in each compiler, out of 653k lines in Clang's front-end path and 78k lines in \tsc's.

\paragraph{Incremental Diagnostics.} 
Both adaptations begin by modifying the lexer to read through a streaming buffer that blocks until the requested characters arrive.
Clang uses essentially one pass---the parser drives the lexer and invokes semantic analysis through callbacks---so pausing the lexer therefore pauses the front-end.

\tsc separates parsing, binding and type checking, so we modify each pass.
The parser creates container nodes when their headers are complete, rather than after their bodies, and emits events when a container starts, a statement completes, or a container ends.
The binder consumes these events and binds child statements when they arrive.
The type-checker checks each statement once it is parsed and bound.

\paragraph{Deferring diagnostics.}
To meet the finality requirements in \S~\ref{sec:design-adapting}, diagnostics that depend on unfinished constructs must be deferred.
Clang already defers such checks, allowing us to reuse its existing behavior.
For \tsc, we delay diagnostics that depend on an open construct until the construct closes.

Our \tsc adaptation relaxes the finality requirement for declaration merging.
A later declaration can merge with a previously checked class, adding members that it appeared to lack.
We do not wait for such declarations, so we may report an error that later code could resolve.
The orchestrator still treats all diagnostics as final, making this exception a source of false rejections.
In validation on real-world repositories, neither compiler falsely accepted a file containing intentionally placed errors.
\tsc falsely rejected 2 of 2,565 files, both due to this exception.

\begin{figure}
    \centering
    \includegraphics[width=\linewidth]{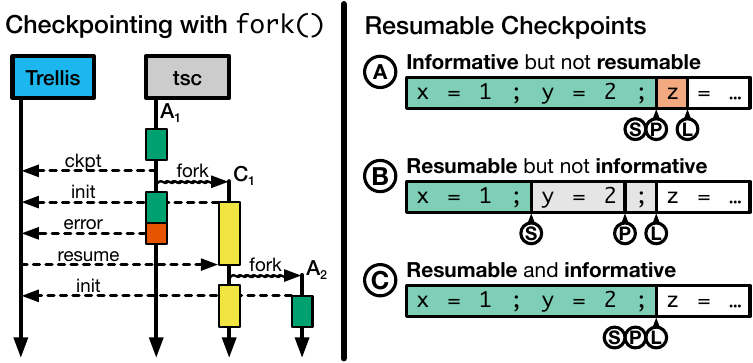}
    \vskip -0.1in
    \caption{Adapting production compilers for \sys.  Left: \texttt{fork()}-based checkpointing.  Right: Resumable and informative checkpoint positions.}
    \label{fig:implementation}
    \vskip -0.1in
\end{figure}

\begin{table*}[t!]
    \caption{Correctness outcomes and resource consumption.  Correctness outcomes are percentages, averaged over seeds and then tasks.  Resource consumption values are geometric means over seeds and then tasks.  Bold indicates the better value in each pair when the difference is statistically significant at $\alpha=0.05$.}
    \label{tab:eval-main}
    \vskip 0.1in
    \begin{center}
    \begin{small}
    \begin{sc}
    \begin{tabular}{llrrrrrrrrrrrr}
        \toprule
        & & \multicolumn{2}{c}{Compiles} & \multicolumn{2}{c}{Builds} & \multicolumn{2}{c}{Tests Pass} & \multicolumn{2}{c}{Resolved} & \multicolumn{2}{c}{Tok (k)} & \multicolumn{2}{c}{Lat (min)} \\
        \cmidrule(lr){3-4} \cmidrule(lr){5-6} \cmidrule(lr){7-8} \cmidrule(lr){9-10} \cmidrule(lr){11-12} \cmidrule(lr){13-14}
        Model & Lang & Base & Hy & Base & Hy & Base & Hy & Base & Hy & Base & Hy & Base & Hy \\
        \midrule
        \multirow{3}{*}{Nano} & All & 84.4 & \textbf{94.9} & 66.2 & \textbf{73.5} & 61.2 & \textbf{68.0} & 11.2 & 12.6 & \textbf{13.2} & 14.0 & \textbf{2.8} & 3.4 \\
         & C/C++ & 82.5 & \textbf{94.2} & 53.1 & \textbf{57.8} & 47.6 & \textbf{51.7} & 8.1 & 9.2 & 15.2 & 16.3 & \textbf{3.1} & 3.6 \\
         & TS & 86.5 & \textbf{95.8} & 80.3 & \textbf{90.3} & 75.8 & \textbf{85.3} & 14.5 & 16.2 & 11.3 & 11.8 & \textbf{2.5} & 3.2 \\
        \midrule
        \multirow{3}{*}{Super} & All & 88.4 & \textbf{93.9} & 71.6 & \textbf{75.1} & 65.8 & \textbf{68.9} & 21.6 & 21.6 & 16.6 & 16.9 & 9.2 & 9.3 \\
         & C/C++ & 86.3 & \textbf{93.6} & 59.9 & 62.1 & 51.0 & 53.5 & 18.4 & 17.8 & 21.1 & 22.0 & 12.0 & 12.7 \\
         & TS & 90.7 & \textbf{94.3} & 84.3 & \textbf{89.2} & 81.8 & \textbf{85.6} & 25.1 & 25.6 & 12.8 & 12.6 & 6.9 & 6.7 \\
        \midrule
        \multirow{3}{*}{Gemma} & All & 96.3 & \textbf{98.9} & 77.8 & \textbf{79.5} & 73.4 & \textbf{75.4} & 31.3 & 32.6 & 9.1 & 9.2 & 8.2 & 8.4 \\
         & C/C++ & 95.7 & \textbf{99.3} & 64.4 & 65.7 & 58.8 & 61.1 & 26.8 & 27.8 & 9.8 & 9.9 & 8.7 & 8.7 \\
         & TS & 96.9 & \textbf{98.6} & 92.3 & \textbf{94.4} & 89.1 & 90.8 & 36.2 & 37.8 & 8.4 & 8.6 & 7.7 & 8.0 \\
        \bottomrule
    \end{tabular}
    \end{sc}
    \end{small}
    \end{center}
    \vskip -0.1in
\end{table*}

\paragraph{Checkpointing with \fork}
Neither Clang nor \tsc exposes serializable checkpoints.
Rather than extract internal state, which may reside on the call stack, we use \fork to preserve the entire process, as illustrated on the left of Figure~\ref{fig:implementation}.

When an active session $A_1$ creates a checkpoint, it informs the orchestrator, invokes \fork, and continues compilation.
Its child $C_1$ registers with the orchestrator and becomes dormant, preserving the compiler state.
When $A_1$ encounters an error, the orchestrator chooses a repair point and asks the closest preceding checkpoint, $C_1$, to resume.
$C_1$ invokes \fork again, creating a new active session $A_2$ that continues from the saved state.
$C_1$ remains dormant and available for subsequent repairs.

Multi-threaded programs generally cannot resume execution safely after \fork.
We therefore run \tsc in the single-threaded QuickJS runtime instead of its default, multi-threaded Node.js runtime.
Checkers that require multi-threaded execution may need a different checkpointing mechanism.

\paragraph{Resumable checkpoints.}
Checkpoints must satisfy the resumability requirement in \S~\ref{sec:design-adapting}.
We also prefer them to be \emph{informative}: checkpointing after an analysis reduces the risk of preserving an undiagnosed error.

The right side of Figure~\ref{fig:implementation} illustrates a challenge in both Clang and \tsc.
To support parser look-ahead, the lexer (L) reads ahead of the parser (P).
Ideally, we would checkpoint after the statement's terminating semicolon is parsed and the statement is semantically analyzed (S).
By this point, however, the lexer has consumed characters beyond the semicolon, meaning that the compiler state depends on characters beyond the checkpoint position (which could be replaced in a repair attempt).
Without the ability to roll back the lexer, this checkpoint is not resumable.

Therefore, we must ensure that the lexer has consumed no characters beyond the reported checkpoint offset at each checkpoint.
Where parser look-ahead is unnecessary, we also reorder compiler logic to analyze code before consuming the next token.
This lets us checkpoint after checking the statement but before reading the suffix that repair may replace.
\section{Evaluation}
\label{sec:evaluation}

\paragraph{Setup.}

We evaluate three recent, open-weight models: Nemotron 3 30B (\emph{Nano})~\cite{nvidia2025nemotron3}, 120B (\emph{Super}), and Gemma 4 31B (\emph{Gemma})~\cite{gemma4}.
Each uses its vendor's recommended sampling parameters with reasoning enabled.
All models use OpenCode~\cite{opencode} as the agent harness and vLLM~\cite{kwon2023pagedattention} as the inference engine, with inference on one NVIDIA H200 GPU.

We combine the C/C++ and TypeScript tasks from SWE-bench Multilingual \cite{jimenez2024swebench} and Multi-SWE-bench~\cite{zan2025multiswebench}, deduplicating tasks by pull request.
We exclude tasks whose gold patch changes no supported files or contain supported files that cannot be checked or fail \sys's checker.
Exclusions include headers that do not compile standalone, repositories whose build flags cannot be extracted, and patches that modify only unsupported file types (for example, several \texttt{ponyc} tasks modify only \texttt{.pony} files).
These criteria exclude 70 of the 537 deduplicated tasks.

We split the remaining tasks into training and test sets, stratifying by repository and gold patch length.
We fit \sys's policy on the training set and evaluate end-to-end results on the test set, which contains 165 C/C++ and 153 TypeScript tasks.

We run each task with five seeds, a two-hour wall-clock limit, and no token limit.
To make evaluation tractable, we run concurrent agent sessions that share a single \sys-serve instance.
We select  concurrency based on KV-cache capacity and use the same concurrency for baseline and \sys runs.
End-to-end latency may therefore reflect contention among sessions.
Token consumption is unaffected except in the 0.1\% of runs that reach the wall-clock limit.

\paragraph{Correctness.} 
Table~\ref{tab:eval-main} reports four correctness metrics: compile rate, build rate, tests passed, and tasks resolved.
Compile rate is the share of files in the final patch that compile independently using the task repository's flags and configuration.
Build rate measures whether the full project builds with the patch.
For C/C++, the build compiles and links hidden tests; for TypeScript, it runs the repository's CI type check.
Builds therefore detect failures beyond \sys's checks during editing, including errors in hidden tests (such as calls to an API missing from the patch), linker errors, errors in unmodified files, and build configuration errors.
Tests passed is the fraction of a task's provided tests that pass; a task is resolved only if all tests pass.
A failed build receives zero for build rate, tests passed, and tasks resolved.

\sys significantly improves compile rate for all three models: by 10.5 percentage points (pp; 12.5\%) for Nano, 5.5 pp (6.2\%) for Super, and 2.7 pp (2.8\%) for Gemma.
These gains extend to full-project builds and functional correctness.
Build rate increases by 7.3 pp (11.0\%) for Nano, 3.5 pp (4.8\%) for Super, and 1.7 pp (2.2\%) for Gemma.
Tests passed increases by 6.7 pp (11\%) for Nano, 3.1 pp (4.8\%) for Super, and 2.0 pp (2.7\%) for Gemma.
All of these improvements are statistically significant at $\alpha=0.05$.

\sys improves task resolution, but not at an $\alpha=0.05$ level.
Task resolution increases by 1.4 pp (12.4\%) for Nano and 1.3 pp (4.0\%) for Gemma and remains unchanged for Super.
Resolving a task requires passing \emph{all} tests, and even the best-performing model, Gemma, resolves fewer than one-third of tasks.
\sys improves functional correctness by reducing compilation failures, but it does not address errors in solutions that compile successfully.

\begin{figure}[t!]
    \centering
    \includegraphics[width=\linewidth]{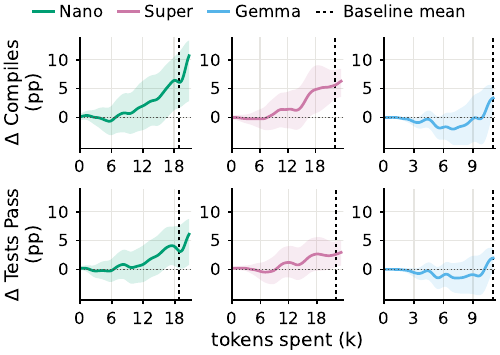}
    \vskip -0.1in
    \caption{Compile rate and tests passed under equal token budgets.  Bands show 95\% confidence intervals.  Dashed lines mark the baseline's mean token consumption; beyond it baseline correctness is held at its final level.  Curves are smoothed for clarity.}
    \label{fig:eval-tokens}
    \vskip -0.1in
\end{figure}

\begin{table}[t!]
    \caption{Time (in ms) per output token, aggregated using geometric means over seeds and then tasks.  Bold indicates the better value in each pair when the difference is statistically significant at $\alpha = 0.05$.}
    \label{tab:eval-latency}
    \vskip 0.1in
    \begin{center}
    \begin{small}
    \begin{sc}
    \begin{tabular}{lrrrrrr}
        \toprule
        & \multicolumn{2}{c}{All} & \multicolumn{2}{c}{C/C++} & \multicolumn{2}{c}{TS} \\
        \cmidrule(lr){2-3} \cmidrule(lr){4-5} \cmidrule(lr){6-7}
        Model & Base & Hy & Base & Hy & Base & Hy \\
        \midrule
        Nano & \textbf{12.8} & 14.7 & \textbf{12.3} & 13.2 & \textbf{13.3} & 16.5 \\
        Super & 33.1 & 33.0 & 34.1 & 34.5 & 32.2 & 31.6 \\
        Gemma & 53.9 & 54.5 & 53.2 & 53.1 & 54.7 & 56.0 \\
        \bottomrule
    \end{tabular}
    \end{sc}
    \end{small}
    \end{center}
    \vskip -0.1in
\end{table}

\paragraph{Resource Consumption.}
Table~\ref{tab:eval-main} also reports output token consumption and latency.
\sys slightly increases both metrics, but these increases are generally not statistically significant at $\alpha=0.05$, except for Nano.

\sys's search and feedback may encourage models to spend more tokens on a task.
Figure~\ref{fig:eval-tokens} isolates its effect on correctness at matched token budgets.
At each budget $B$, attempts that exceed $B$ receive a score of zero and are charged $B$ tokens; attempts that finish within $B$ retain their scores and are charged their actual token consumption.
Varying $B$ traces a curve of mean correctness against mean token consumption.

The results suggest that increased token consumption alone does not explain \sys's correctness gains.
For Nano and Super, \sys improves correctness at nearly all levels of token consumption.
For Gemma, correctness is initially lower at small budgets but exceeds the baseline before reaching the baseline's mean token consumption.

Table~\ref{tab:eval-latency} reports \sys's effect on time per output token (TPOT).
\sys does not significantly increase TPOT for Super or Gemma, suggesting that their higher overall latency primarily reflects increased token consumption.
The exception is Nano, for which \sys significantly increases both TPOT and overall latency.
This is because Nano is the smallest model and generates tokens quickly enough that parallel execution cannot fully hide compiler latency.
The effect is larger for \tsc, which takes longer to produce a final verdict.

\begin{table}[t!]
    \caption{Correctness outcomes and resource consumption for Nano with and without search.}
    \label{tab:eval-search}
    \vskip 0.1in
    \begin{center}
    \begin{small}
    \begin{sc}
    \begin{tabular}{lrr}
    \toprule
     & Feedback & Feedback + Search \\
    \midrule
    Compiles & 73.9 & \textbf{81.8} \\
    Builds & 52.1 & \textbf{57.4} \\
    Tests Pass & 49.4 & 52.3 \\
    Resolved & 12.5 & 11.7 \\
    \midrule
    Agent tokens & 12{,}926 & 13{,}522 \\
    Search tokens & 0 & 1{,}435 \\
    Latency (s) & \textbf{151} & 199 \\
    \bottomrule
    \end{tabular}
    \end{sc}
    \end{small}
    \end{center}
    \vskip -0.1in
\end{table}

\paragraph{Search Ablation.}

Table~\ref{tab:eval-search} evaluates the effect of search.
We select points in Nano's training traces where search was initiated and resume each point five times with search enabled and five times with search disabled.
Search increases compile rate by 7.9 pp, build rate by 5.3 pp, and tests passed by 2.9 pp.

Search also increases token consumption, primarily through additional search tokens.
When \sys abandons a search, the agent must still repair the error, making those search tokens pure overhead.
Compilation overhead also contributes to the increase in latency, as discussed above.

\begin{figure}[t!]
    \centering
    \includegraphics[width=\linewidth]{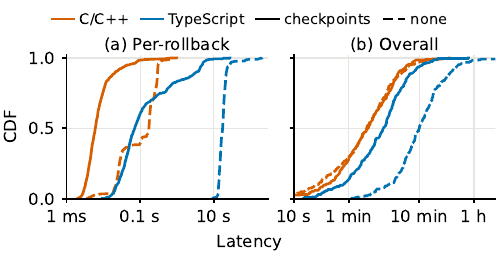}
    \vskip -0.1in
    \caption{Latency with and without checkpointing.}
    \label{fig:eval-checkpointing}
    \vskip -0.1in
\end{figure}

\paragraph{Checkpointing Ablation.}
Figure~\ref{fig:eval-checkpointing} shows how checkpointing affects per-rollback and end-to-end latency.
For the left panel, we replay recorded rollouts from Nano's training traces and measure the time to final verdict.
Checkpointing reduces latency by one order of magnitude for Clang and two orders of magnitude for \tsc.
The right panel shows that Clang's improvement does not translate into lower end-to-end latency: Clang runs faster than token generation and can catch up  even when restarting from scratch.

For TypeScript, checkpointing reduces end-to-end latency by an order of magnitude.
In addition to processing tokens more slowly, \tsc incurs substantial overhead constructing TypeScript Programs, the compiler objects used to parse and bind multiple files.
Checkpointing allows these objects to be reused.
\section{Discussion}
\label{sec:discussion}

\paragraph{Model capability.}
We evaluated several mid-size models.
While improving their performance is valuable in its own right, a natural question is whether \sys's benefits extend to larger, more capable models.
Frontier models may make fewer errors in our evaluation setting, but their errors are more computationally expensive to repair.
When no errors occur, \sys incurs no additional token cost and negligible latency overhead, as shown in \S~\ref{sec:evaluation}.
\sys's design could also support less common programming languages and more complex static analyses, settings that may challenge even highly capable models while enabling \sys to enforce stronger guarantees.

\paragraph{Language features and diagnostic latency.}
\S\ref{sec:implementation} illustrates how language features can delay diagnostics.
A later declaration can add members to a previously checked class, resolving an earlier missing-member error.
Our implementation leaves these diagnostics undeferred, accepting rare false rejections to reduce diagnostic latency.

This example suggests a broader trade-off between language flexibility and the strength and timeliness of incremental diagnostics.
Which language restrictions enable earlier, stronger diagnostics without unduly limiting the programs agents can produce?
We focus on efficient incremental checking using existing models and compilers, leaving the exploration of such restrictions to future work.

\paragraph{\sys-aware agents.}
Our incremental compilers can produce executable output---machine code or JavaScript---in addition to front-end diagnostics.
With appropriate instructions and tool interfaces, agents could reuse this output and avoid a separate compilation tool call.
We leave this opportunity to future work, as our focus is on transparent integration with existing models and coding agents.
\section{Related Work}
\label{sec:related-work}

\paragraph{Program repair and coding agents.}
LLMs can revise generated code based on execution or compiler feedback~\cite{chen2024teaching, olausson2024selfrepair, bi2024iterative}. 
Coding agents incorporate checking and repair into tool-driven workflows~\cite{yang2024sweagent, openhands, bouzenia2025repairagent}, and some automatically lint completed edits~\cite{yang2024sweagent, aider2026aider}.

Agent-directed repair relies on the model to request checks, interpret diagnostics, and issue revisions through subsequent tool calls.
\sys directly manages localized retries during generation, while returning control when broader agent actions are needed.

\paragraph{Guiding code generation.}
Grammar-constrained decoding restricts token selection to enforce syntactic validity~\cite{koo2024automata, dong2024xgrammar, guidanceai2026llguidance, park2025grammarconstrained}. 
Constrained semantic decoding extends this approach to semantic properties~\cite{poesia2022synchromesh, nagy2026chopchop, mundler2025typeconstrained, li2025correctness}.
Other approaches query language servers or completion engines~\cite{agrawal2023monitorguided, wei2023copiloting, blinn2024statically}, or complete partial programs for standard compilers to consume~\cite{jiang2025rocode, mundler2026generative}.
These approaches require specialized parsers and must take care to preserve the compiler's full checking behavior.
\sys instead integrates production compilers through a streaming interface, allowing sound diagnostics to arrive and preserving all production compiler analyses.

Several systems roll back and regenerate code when errors are detected~\cite{jiang2025rocode, wang2025semguard, li2025adaptrack}.
These papers use specialized decision criteria, like model entropy or a learned evaluator.
\sys's policy interface supports these search strategies and improves efficiency by reusing prior computation for later repair attempts.
\section{Conclusion}
\label{sec:conclusion}

We presented \sys, a system that integrates compiler-guided repair into agentic code generation.
\sys adapts production compilers to provide incremental feedback, overlaps checking with generation, and uses rollback-based search and reusable checkpoints to make localized repair efficient.
It accommodates intermediate errors and returns control when broader agent-directed changes are needed.
Integrated into OpenCode, \sys improves compilation rate by up to 13\% and test pass rate by up to 11\% on C/C++ and TypeScript software engineering tasks, with comparable token consumption.
These results demonstrate that compiler feedback can support efficient repair within generation turns while preserving the agent's ability to complete changes across edits and files. \sys's source code will be publicly available.

\newpage

\bibliography{confs_long,refs}

@string{ACL=proc # "Annual Meeting of the Association for Computational Linguistics (ACL)"}

@string{ASE=proc # "IEEE/ACM International Conference on Automated Software Engineering (ASE)" }

@string{COLM=proc # "Conference on Language Modeling"}

@string{FSE=proc # "ACM International Conference on the Foundations of Software Engineering (FSE)"}

@string{ICLR=proc # "International Conference on Learning Representations (ICLR)"}

@string{ICML=proc # "International Conference on Machine Learning (ICML)"}

@string{ICSE=proc # "International Conference on Software Engineering (ICSE)"}

@string{MLSYS=proc # "Conference on Machine Learning and Systems (MLSys)"}

@string{NEURIPS=proc # "Conference on Neural Information Processing Systems (NeurIPS)"}

@string{OOPSLA=proc # "Object-oriented Programming, Systems, Languages, and Applications (OOPSLA)"}

@string{PLDI=proc # "ACM SIGPLAN Conference on Programming Language Design and Implementation (PLDI)"}

@string{POPL=proc # "ACM SIGPLAN-SIGACT Symposium on Principles of Programming Languages (POPL)"}

@string{SOSP=proc # "ACM Symposium on Operating Systems Principles (SOSP)"}

@misc{aider2026aider,
    author = {{Aider-AI}},
    title = {{Aider}: {AI} pair programming in your terminal},
    year = {2026},
    url = {https://aider.chat},
    note = {Accessed: 2026-04-09}
}

@misc{anthropic2026claudecode,
    author = {{Anthropic}},
    title = {{Claude Code}},
    year = {2026},
    url = {https://claude.com/product/claude-code},
    note = {Accessed: 2026-04-09}
}

@misc{opencode,
    author = {{OpenCode}},
    title = {{The open source AI coding agent}},
    year = {2026},
    url = {https://opencode.ai/},
    note = {Accessed: 2026-10-29}
}

@inproceedings{openhands,
    author = {Wang, Xingyao and Li, Boxuan and Song, Yufan and Xu, Frank F. and Tang, Xiangru and Zhuge, Mingchen and Pan, Jiayi and Song, Yueqi and Li, Bowen and Singh, Jaskirat and others},
    title = {{OpenHands}: an open platform for {AI} software developers as generalist agents},
    booktitle = ICLR,
    year = {2025},
    url = {https://openreview.net/forum?id=OJd3ayDDoF}
}

@misc{openai2026codex,
    author = {{OpenAI}},
    title = {{Codex}},
    year = {2026},
    url = {https://openai.com/codex/},
    note = {Accessed: 2026-04-09}
}

@misc{llvm2026clang,
    author = {{LLVM Project}},
    title = {{Clang}: a {C} language family frontend for {LLVM}},
    year = {2026},
    url = {https://clang.llvm.org},
    note = {Accessed: 2026-04-09}
}

@inproceedings{kwon2023pagedattention,
    author = {Kwon, Woosuk and Li, Zhuohan and Zhuang, Siyuan and Sheng, Ying and Zheng, Lianmin and Yu, Cody Hao and Gonzalez, Joseph and Zhang, Hao and Stoica, Ion},
    title = {Efficient memory management for large language model serving with {PagedAttention}},
    booktitle = SOSP,
    year = {2023},
    doi = {10.1145/3600006.3613165}
}

@inproceedings{bi2024iterative,
    author = {Bi, Zhangqian and Wan, Yao and Wang, Zheng and Zhang, Hongyu and Guan, Batu and Lu, Fangxin and Zhang, Zili and Sui, Yulei and Jin, Hai and Shi, Xuanhua},
    title = {Iterative refinement of project-level code context for precise code generation with compiler feedback},
    booktitle = {Findings of the Association for Computational Linguistics: ACL 2024},
    year = {2024},
    month = aug,
    doi = {10.18653/v1/2024.findings-acl.138}
}

@inproceedings{bouzenia2025repairagent,
    author = {Bouzenia, Islem and Devanbu, Premkumar and Pradel, Michael},
    title = {{RepairAgent}: an autonomous, {LLM}-based agent for program repair},
    booktitle = ICSE,
    year = {2025},
    doi = {10.1109/ICSE55347.2025.00157}
}

@inproceedings{chen2024teaching,
    author = {Chen, Xinyun and Lin, Maxwell and Sch{\"a}rli, Nathanael and Zhou, Denny},
    title = {Teaching large language models to self-debug},
    booktitle = ICLR,
    year = {2024},
    url = {https://openreview.net/forum?id=KuPixIqPiq}
}

@inproceedings{olausson2024selfrepair,
    author = {Olausson, Theo X. and Inala, Jeevana Priya and Wang, Chenglong and Gao, Jianfeng and Solar-Lezama, Armando},
    title = {Is self-repair a silver bullet for code generation?},
    booktitle = ICLR,
    year = {2024},
    url = {https://openreview.net/forum?id=y0GJXRungR}
}

@inproceedings{yang2024sweagent,
    author = {Yang, John and Jimenez, Carlos E. and Wettig, Alexander and Lieret, Kilian and Yao, Shunyu and Narasimhan, Karthik and Press, Ofir},
    title = {{SWE-agent}: agent-computer interfaces enable automated software engineering},
    booktitle = NEURIPS,
    year = {2024},
    url = {https://openreview.net/forum?id=mXpq6ut8J3}
}

@inproceedings{dong2024xgrammar,
    author = {Dong, Yixin and Ruan, Charlie F and Cai, Yaxing and Lai, Ruihang and Xu, Ziyi and Zhao, Yilong and Chen, Tianqi},
    title = {{XGrammar}: flexible and efficient structured generation engine for large language models},
    booktitle = MLSYS,
    year = {2025},
    url = {https://openreview.net/forum?id=rjQfX0YgDl}
}

@misc{guidanceai2026llguidance,
    author = {{guidance-ai}},
    title = {Low-level guidance (llguidance)},
    year = {2026},
    url = {https://github.com/guidance-ai/llguidance},
    note = {Accessed: 2026-04-09}
}

@inproceedings{koo2024automata,
    author = {Koo, Terry and Liu, Frederick and He, Luheng},
    title = {Automata-based constraints for language model decoding},
    booktitle = COLM,
    year = {2024},
    url = {https://openreview.net/forum?id=BDBdblmyzY}
}

@inproceedings{park2025grammarconstrained,
    author = {Park, Kanghee and Zhou, Timothy and D'Antoni, Loris},
    title = {Flexible and efficient grammar-constrained decoding},
    booktitle = ICML,
    year = {2025},
    url = {https://openreview.net/forum?id=L6CYAzpO1k}
}

@inproceedings{li2025correctness,
    author = {Li, Lingxiao and Rahili, Salar and Zhao, Yiwei},
    title = {Correctness-guaranteed code generation via constrained decoding},
    booktitle = COLM,
    year = {2025},
    url = {https://openreview.net/forum?id=CYiXNIQegF}
}

@inproceedings{mundler2025typeconstrained,
    author = {M{\"u}ndler, Niels and He, Jingxuan and Wang, Hao and Sen, Koushik and Song, Dawn and Vechev, Martin},
    title = {Type-constrained code generation with language models},
    booktitle = PLDI,
    year = {2025},
    doi = {10.1145/3729274}
}

@inproceedings{nagy2026chopchop,
    author = {Nagy, Shaan and Zhou, Timothy and Polikarpova, Nadia and D'Antoni, Loris},
    title = {{ChopChop}: a programmable framework for semantically constraining the output of language models},
    booktitle = POPL,
    year = {2026},
    doi = {10.1145/3776708}
}

@inproceedings{poesia2022synchromesh,
    author = {Poesia, Gabriel and Polozov, Alex and Le, Vu and Tiwari, Ashish and Soares, Gustavo and Meek, Christopher and Gulwani, Sumit},
    title = {{Synchromesh}: reliable code generation from pre-trained language models},
    booktitle = ICLR,
    year = {2022},
    url = {https://openreview.net/forum?id=KmtVD97J43e}
}

@inproceedings{agrawal2023monitorguided,
    author = {Agrawal, Lakshya and Kanade, Aditya and Goyal, Navin and Lahiri, Shuvendu K and Rajamani, Sriram},
    title = {Monitor-guided decoding of code {LM}s with static analysis of repository context},
    booktitle = NEURIPS,
    year = {2023},
    url = {https://openreview.net/forum?id=qPUbKxKvXq}
}

@inproceedings{blinn2024statically,
    author = {Blinn, Andrew and Li, Xiang and Kim, June Hyung and Omar, Cyrus},
    title = {Statically contextualizing large language models with typed holes},
    booktitle = OOPSLA,
    year = {2024},
    doi = {10.1145/3689728}
}

@inproceedings{jiang2025rocode,
    author = {Jiang, Xue and Dong, Yihong and Tao, Yongding and Liu, Huanyu and Jin, Zhi and Li, Ge},
    title = {{ROCODE}: integrating backtracking mechanism and program analysis in large language models for code generation},
    booktitle = ICSE,
    year = {2025},
    doi = {10.1109/ICSE55347.2025.00133}
}

@article{li2025adaptrack,
    author = {Li, Yongmin and Li, Jia and Li, Ge and Jin, Zhi},
    title = {{AdapTrack}: constrained decoding without distorting {LLM}'s output intent},
    journal = {arXiv preprint arXiv:2510.17376},
    year = {2025},
    url = {https://arxiv.org/abs/2510.17376}
}

@inproceedings{wang2025semguard,
    author = {Wang, Qinglin and Sun, Zhihong and Wang, Ruyun and Huang, Tao and Jin, Zhi and Li, Ge and Lyu, Chen},
    title = {{SemGuard}: real-time semantic evaluator for correcting {LLM}-generated code},
    booktitle = ASE,
    year = {2025},
    doi = {10.1109/ASE63991.2025.00160}
}

@inproceedings{wei2023copiloting,
    author = {Wei, Yuxiang and Xia, Chunqiu Steven and Zhang, Lingming},
    title = {Copiloting the copilots: fusing large language models with completion engines for automated program repair},
    booktitle = FSE,
    year = {2023},
    doi = {10.1145/3611643.3616271}
}

@inproceedings{zan2025multiswebench,
    author = {Zan, Daoguang and Huang, Zhirong and Liu, Wei and Chen, Hanwu and Xin, Shulin and Zhang, Linhao and Liu, Qi and Li, Aoyan and Chen, Lu and Zhong, Xiaojian and Liu, Siyao and Xiao, Yongsheng and Chen, Liangqiang and Zhang, Yuyu and Su, Jing and Liu, Tianyu and Long, Rui and Ding, Ming and Xiang, Liang},
    title = {{Multi-SWE-bench}: a multilingual benchmark for issue resolving},
    booktitle = NEURIPS,
    year = {2025},
    note = {Datasets and Benchmarks Track},
    url = {https://arxiv.org/abs/2504.02605}
}

@article{bellman1954theory,
  author  = {Bellman, Richard},
  title   = {The Theory of Dynamic Programming},
  journal = {Bulletin of the American Mathematical Society},
  year    = {1954},
  volume  = {60},
  number  = {6},
  pages   = {503--515},
  doi     = {10.1090/S0002-9904-1954-09848-8}
}

@inproceedings{jimenez2024swebench,
    author = {Jimenez, Carlos E. and Yang, John and Wettig, Alexander and Yao, Shunyu and Pei, Kexin and Press, Ofir and Narasimhan, Karthik},
    title = {{SWE-bench}: can language models resolve real-world {GitHub} issues?},
    booktitle = ICLR,
    year = {2024},
    url = {https://openreview.net/forum?id=VTF8yNQM66}
}

@misc{microsoft2026typescript,
    author = {{Microsoft}},
    title = {{TypeScript}},
    year = {2026},
    url = {https://www.typescriptlang.org/},
    note = {Accessed: 2026-10-29}
}

@inproceedings{selfconditioning,
    author = {Sinha, Akshit and Arun, Arvindh and Goel, Shashwat and Staab, Steffen and Geiping, Jonas},
    title = {The illusion of diminishing returns: measuring long horizon execution in {LLM}s},
    booktitle = ICLR,
    year = {2026},
    pages = {35184--35211}
}

@inproceedings{lostmultiturn,
    author = {Laban, Philippe and Hayashi, Hiroaki and Zhou, Yingbo and Neville, Jennifer},
    title = {{LLM}s get lost in multi-turn conversation},
    booktitle = ICLR,
    year = {2026},
    pages = {54738--54778}
}

@article{agentdiet,
    author = {Xiao, Yuan-An and Gao, Pengfei and Peng, Chao and Xiong, Yingfei},
    title = {Reducing cost of {LLM} agents with trajectory reduction},
    journal = {Proceedings of the ACM on Software Engineering},
    year = {2026},
    volume = {3},
    number = {FSE},
    doi = {10.1145/3797084}
}

@article{nvidia2025nemotron3,
    author = {{NVIDIA}},
    title = {{NVIDIA Nemotron 3}: efficient and open intelligence},
    journal = {arXiv preprint arXiv:2512.20856},
    year = {2025},
    url = {https://arxiv.org/abs/2512.20856}
}

@misc{gemma4,
    author = {{Gemma Team, Google DeepMind}},
    title = {{Gemma} 4},
    year = {2026},
    url = {https://huggingface.co/google/gemma-4-31B-it},
    note = {Accessed: 2026-10-29}
}

@article{mundler2026generative,
    author = {M{\"u}ndler-Sasahara, Niels and Venev, Hristo and Song, Dawn and Vechev, Martin and He, Jingxuan},
    title = {Generative compilation: on-the-fly compiler feedback as {AI} generates code},
    journal = {arXiv preprint arXiv:2607.13921},
    year = {2026},
    url = {https://arxiv.org/abs/2607.13921}
}

@article{farewell1982mixture,
  author  = {Farewell, V. T.},
  title   = {The Use of Mixture Models for the Analysis of Survival Data with Long-Term Survivors},
  journal = {Biometrics},
  year    = {1982},
  volume  = {38},
  number  = {4},
  pages   = {1041--1046},
  doi     = {10.2307/2529885}
}

@article{weitzman1979optimal,
  author  = {Weitzman, Martin L.},
  title   = {Optimal Search for the Best Alternative},
  journal = {Econometrica},
  year    = {1979},
  volume  = {47},
  number  = {3},
  pages   = {641--654},
  doi     = {10.2307/1910412}
}

@phdthesis{duff2002optimal,
  author = {Duff, Michael O'Gordon},
  title  = {Optimal Learning: Computational Procedures for {B}ayes-Adaptive {M}arkov Decision Processes},
  school = {University of Massachusetts Amherst},
  year   = {2002},
  url    = {https://moduff.github.io/DuffThesis.pdf}
}

@article{teh2006hierarchical,
  author  = {Teh, Yee Whye and Jordan, Michael I. and Beal, Matthew J. and Blei, David M.},
  title   = {Hierarchical {D}irichlet Processes},
  journal = {Journal of the American Statistical Association},
  year    = {2006},
  volume  = {101},
  number  = {476},
  pages   = {1566--1581},
  doi     = {10.1198/016214506000000302}
}
\bibliographystyle{mlsys2025style/mlsys2025}

\section*{AI Usage Disclosure}

We used AI tools during the implementation, evaluation, and writing of this paper.
The detailed usage descriptions are as follows:
\emph{Code:} We used coding agents to assist in the implementation of \sys.  The core design is the authors'.
\emph{Experiments:} We used AI assistants to generate test scripts and organize experiment runs.  The authors designed all experiments.
\emph{Literature search:} We used AI assistants to help search related work.
\emph{Writing:} We planned and drafted every section of the paper, but used AI assistants to edit and revise parts of the text.

\newpage

\appendix

\section{Search Policy}
\label{app:policy}

At each decision, the search has a set $B$ of candidate boundaries.  The policy $\pi$ either retries from one of these boundaries or gives up.

\paragraph{Success model.}
The probability that a retry from $b$ is accepted is a logistic regression with random effects:
\begin{equation}
  p_b = \mathbb{E}\big[\,\sigma(\beta^\top x + \gamma^\top z_b + u + v_b)\,\big],
\end{equation}
where $x$ holds search-level features, $z_b$ boundary-level features (Table~\ref{tab:policy:terms}), $u \sim \mathcal{N}(0, \sigma_u^2)$ is a search effect, and $v_b \sim \mathcal{N}(0, \sigma_v^2)$ a boundary effect.
The expectation is over the posterior of $(u, v)$ given the failed retries $\mathcal{F}$:
\begin{equation}
  p(u, v \mid \mathcal{F}) \propto \mathcal{N}(u)\, \mathcal{N}(v) \prod_{i \in \mathcal{F}} \big(1 - p_i(u, v)\big).
\end{equation}
A failure at $b$ lowers $v_b$, which affects only $b$, and $u$, which affects every boundary.
For one model (Nano), the regression also includes a mixture component for searches that cannot succeed~\cite{farewell1982mixture}.

\paragraph{Cost model.}
The expected token cost of a retry from $b$ is
\begin{equation}
  c_b = p_b\, T^{s}_b + (1 - p_b)\, T^{f}_b,
\end{equation}
where $T^{s}_b$ and $T^{f}_b$ are the expected tokens of a successful and a failed retry.
Both are log-normal regressions on the same features, with search and boundary effects updated after each failure.

\paragraph{Give-up cost.}
Giving up costs $H$, the agent's expected tokens to repair the error itself.
$H$ is a regression on features of the error, fitted on repair windows from agent traces, with a smearing correction.

\paragraph{Error transitions.}
A failed retry often produces a new error, which adds new boundaries.
The next error $e$ after a failure from $b$ follows a hierarchical Chinese restaurant process~\cite{teh2006hierarchical}:
\begin{equation}
  P(e \mid b) \propto n_{b,e} + \alpha_b\, P_0(e),
\end{equation}
where $n_{b,e}$ counts earlier failures from $b$ that ended in $e$, and the base measure $P_0$ is shared across the search.

\paragraph{Decision rule.}
The policy minimizes expected tokens in a Bayes-adaptive Markov decision process~\cite{duff2002optimal}:
\begin{align}
  Q(b) &= c_b + (1 - p_b)\, \mathbb{E}\big[V(s')\big], \\
  V(s) &= \min\big\{H,\ \min_b Q(b)\big\},
\end{align}
where $s'$ is the belief state after a failure at $b$.
$\pi$ gives up if $\min_b Q(b) \ge H$, and otherwise retries from $\arg\min_b Q(b)$.
We solve the equation with a two-step lookahead.
At the leaves, each boundary is retried until its index $c_b / p_b$ exceeds $H$, as in reservation-value search~\cite{weitzman1979optimal}.

\paragraph{Fitting.}
All parameters are fitted on the training set.
The data are retries resampled offline from recorded error states, plus live \sys{} traces.
Each regression is fitted with a group lasso, where each group is a feature: a categorical feature's levels together, a continuous feature with its square.
The penalty is chosen by task-level cross-validation, and the selected features are refitted with a ridge penalty.
We alternate live training runs with refits, for four rounds per model.

\begin{table}[t]
  \centering
  \small
  \caption{Features supplied to the success, cost and transition regressions. Continuous features enter as standardized $\log(1+x)$ and its square; counts as $\log(1+n)$. $^\ast$Kinds: refusal, runaway, empty.}
  \label{tab:policy:terms}
  \vskip 0.1in
  \sc
  \begin{tabular}{@{}lp{0.58\linewidth}@{}}
    \toprule
    Feature & Description \\
    \midrule
    \multicolumn{2}{@{}l}{Search level ($x$)} \\
    language & C, C++ or TypeScript \\
    mode & region edit or whole file \\
    file size & characters \\
    failures & failed retries so far \\
    fail kinds & failures by kind$^\ast$ \\
    \midrule
    \multicolumn{2}{@{}l}{Boundary level ($z_b$)} \\
    error class & syntax, name, type, other \\
    error code & compiler diagnostic \\
    unseen name & names an identifier absent before it \\
    error before & lies before the edited region \\
    relation & structural relation of $b$ to the error \\
    rank & position of $b$ back from the error \\
    kind & boundary type \\
    distance & bytes from $b$ to the error \\
    remainder & bytes from $b$ to the end \\
    local fails & failed retries from $b$ \\
    local kinds & failures from $b$ by kind \\
    \bottomrule
  \end{tabular}
  \vskip -0.1in
\end{table}

\end{document}